\documentclass[12pt]{article}
\usepackage{amsfonts}
\usepackage{amsmath}
\usepackage{latexsym}
\usepackage{amssymb}
\usepackage[mathcal]{euscript}
\usepackage{amscd}

\def\ben{\begin{equation}}
\def\een{\end{equation}}
\def\bea{\begin{eqnarray}}
\def\eea{\end{eqnarray}}
\input amssym.def
\input amssym.tex
\def\bea{\begin{eqnarray}}
\def\eea{\end{eqnarray}}
\input amssym.def
\input amssym.tex
\input{tcilatex}
\begin{document}

\title{Maximum Tension: with and without a cosmological constant}
\author{John D. Barrow$^{1}$ and G. W. Gibbons$^{1,2,3,4}$ \\
$^{1}$DAMTP, Centre for Mathematical Sciences,\\
Wilberforce Rd., Cambridge CB3 0WA, UK\\
University of Cambridge, U.K.,\\
$^{2}$LE STUDIUM, Loire Valley Institute for Advanced Studies,\\
Tours and Orleans, 45000 Orleans France \\
$^{3}$Laboratoire de Math\'{e}matiques et de Physique \\
Th\'{e}orique, Universit\'{e} de Tours, 37200 Tours, France \\
$^{4}$Dept. of Physics and Astronomy,\\
Univ. Pennsylvania, Philadelphia, PA 19104-6396, USA\\
}
\date{\today }
\maketitle

\begin{abstract}
We discuss various examples and ramifications of the conjecture that there
exists a maximum force (or tension) in general relativistic systems. We
contrast this situation with that in Newtonian gravity, where no maximum
force exists, and relate it to the existence of natural units defined by
constants of Nature and the fact that the Planck units of force and power do
not depend on Planck's constant. We discuss how these results change in
higher dimensions where the Planck units of force are no longer non-quantum.
We discuss the changes that might occur to the conjecture if a positive
cosmological constant exists and derive a maximum force bound using the
Kottler-Schwarzschild-de Sitter black hole.
\end{abstract}

\section{Introduction}

Some time ago it was conjectured (Gibbons 2002) that in general relativity
there should be a maximum value to any physically attainable force (or
tension) given by 
\begin{equation}
F_{\max }=\frac{c^{4}}{4G}\,,  \label{1}
\end{equation}%
where $c$ is the velocity of light and $G$ is the Newtonian gravitational
constant. The sharp factor of $1/4$ is supported by considering the maximum
deficit angle of a cosmic string (Gibbons, 2002). This gives rise to the
closely related conjecture that there is a maximum power defined by

\begin{equation}
P_{\max }=cF_{\max }=\frac{c^{5}}{4G},  \label{2}
\end{equation}%
the so-called Dyson Luminosity\textit{\ }(Dyson, 1963),\textit{\ }or some
multiple of it to account for geometrical factors $O(1)$. This would be the
maximum possible luminosity in gravitational waves, or indeed other forms of
radiation that an isolated system may emit (Schiller 1997, 2006, Sperhake et
al 2013, Cardoso 2013). Schiller (1997, 2006) has come to the same
conclusion and proposed a stronger thesis: that the existence of a maximum
force implies general relativity, just as a maximum velocity characterises
special relativity. This claim is much less clear since it requires in
effect a proof of cosmic censorship. It is also necessary to choose quite
subtle energy conditions in order to avoid the formation of sudden
singularities (Barrow, 2004) where unbounded pressure forces will occur. As
we will see below, it would probably be specific to $3+1$ dimensional
spacetime.

These conjectures provoke a number of comments and a discussion of some
further ramifications of bounded forces in gravitating systems.

\subsection{Newtonian gravity}

First, there is no corresponding principle of maximum force in Newtonian
gravity. Point particle masses can get arbitrarily close to one another and
so the forces between them are unbounded in principle. An important example
was constructed by Xia (1992). He considered a 5-body system consisting of
two counter-rotating binaries systems of equal mass points with zero net
angular momentum, between which a lighter point mass oscillates back and
forth along the line joining the mass centres of the two binaries. For a
Cantor set of initial conditions, the system expands to infinite size in
finite time and the lighter particle undergoes an infinite number of
oscillations during that period. This totally unexpected behaviour is
facilitated by the arbitrarily large gravitational forces that are possible
between the point particles as their separations tend to zero and the
absence of any speed limit for information transmission. This unusual
Newtonian behaviour has no general relativistic counterpart: two particles
of mass $M$ whose centres approach closer than $d=4GM/c^{2}$ will find
themselves inside a black hole horizon. This is a simple form of 'cosmic
censorship' whereby horizon formation prevents the effects of an arbitrary
strong infinite force being visible.

\subsection{Cosmic strings}

Interesting circumstantial evidence for the maximum force conjecture can be
seen in the general-relativistic metrics for line sources (Marder, 1959) or
cosmic strings (Vilenkin, 1981). A static source with $\rho
+\sum_{i=1}^{3}p_{i}=0$ for its density, $\rho $, and principal pressures, $%
p_{i}$, which has a constant mass per unit length$,$ $\mu ,$ has no
Newtonian gravitational source ($\nabla ^{2}\Phi _{N}=0$ for the Newtonian
potential, $\Phi _{N}$). Yet, it supports a conical metric which is flat
spacetime with a missing wedge angle $\Delta \theta =8\pi G\mu c^{-2}$ which
will exceed $2\pi $ and encompass the entire spacetime if $F>F_{\max }$.
Thus the maximum force conjecture is linked to the structure of static
cosmic strings even though they exert no forces on themselves or other
particles. This example supports the correctness of the $1/4$ factor in eq.(%
\ref{1}), noted in Gibbons (2002.. The dimensionless factor in eq.(\ref{2})
does not seem to be so precisely determined at present due to possible
dependence on geometrical factors associated with power generating
configurations involving many bodies.

\subsection{Natural units}

The maximum force conjecture also shows how dimensional analysis can still
provide fundamental insights. If we seek to construct natural units for
various physical quantities from the constants $c,G$ and $\hslash $ then we
can form basic Planck units (Planck, 1899) of mass, length, and time in the
usual way:

\begin{eqnarray}
L_{pl} &=&\left( \frac{G\hslash }{c^{3}}\right) ^{1/2},  \label{3} \\
T_{pl} &=&\left( \frac{G\hslash }{c^{5}}\right) ^{1/2},  \label{4} \\
M_{pl} &=&\left( \frac{\hbar c}{G}\right) ^{1/2}.  \label{5}
\end{eqnarray}

\bigskip

These examples of a fundamental length, mass and time all contain $G,c$ and $%
\hslash $ and so have a quantum significance. However, there are associated
quantities, like the Planck force, $F_{pl}=c^{4}/G,$ and power $%
P_{pl}=c^{5}/G$, that do not contain $\hslash $ and so are entirely
classical. Whenever Planck units can be found for a quantity that does not
contain $\hslash ,$ this implies that it plays a fundamental role in
classical gravity.

Planck's units of 1899 (Planck, 1899) were not the first set of natural
units to be proposed (see Barrow and Tipler, 1986). In 1881, Johnson Stoney
proposed a set of fundamental units involving $c,G$ and $e$ (Stoney, 1881,
Barrow 1983, 2002). This was far sighted in that $c$ did not yet have its
fundamental relativistic significance and the electron, whose existence and
charge $e$ were predicted by Stoney in 1874 was not discovered by Thomson
and colleagues until 1897 (Thomson, 1879). Stoney's natural units were

\begin{eqnarray}
L_{S} &=&\left( \frac{Ge^{2}}{c^{4}}\right) ^{1/2},  \label{6} \\
T_{S} &=&\left( \frac{Ge^{2}}{c^{6}}\right) ^{1/2},  \label{7} \\
M_{S} &=&\left( \frac{e^{2}}{G}\right) ^{1/2}.  \label{8}
\end{eqnarray}%
The fine structure constant $e^{2}/\hslash c$ can be used to convert between
Stoney and Planck units and their magnitudes just differ by approximately$%
\sqrt{137},$ (Barrow, 2002).

\subsection{Higher dimensions}

In $N$ space dimensions there are generalisations of these natural units. By
Gauss's theorem we have $[G]=M^{-1}L^{N}T^{-2}$ and $[e^{2}]=ML^{N}T^{-2},$
but $[c]=LT^{-1}$ and $[\hslash ]=ML^{2}T^{-1}$ as before. hence the
dimensionless quantity that generalises the fine structure constant when $%
N=3 $ to arbitrary dimensions is (Barrow and Tipler, 1986)

\begin{equation}
\hslash ^{2-N}e^{N-1}G^{(3-N)/2}c^{N-4}.  \label{9}
\end{equation}%
Only when $N=3$ is gravity excluded. If we confine attention to combinations
of $G,c$ and $\hslash $ then we find that in $N$ dimensions the physical
quantity that does not include a dependence on $\hslash $ is only a force
(or power) when $N=2.$ For general $N,$ the fundamental non-quantum quantity
is

\begin{equation}
Q\equiv mass\times (acceleration)^{N-2}.  \label{10}
\end{equation}

This reduces to force in three dimensions and suggests a generalised
conjecture that in $N$-dimensional general relativity there will be an upper
bound on the magnitude of $Q$ determined by

\begin{equation}
Q_{\max }=\frac{c^{2(N-1)}}{G}.  \label{11}
\end{equation}%
A calculation using the $N$-dimensional Schwarzschild metric gives the
dimensionless factor. The horizon radius of the $N$-dimensional
Schwarzschild metric is

\begin{equation}
r=\left[ \frac{16\pi GM}{(N-1)\Omega _{N-1}c^{2}}\right] ^{\frac{1}{N-2}},
\label{11a}
\end{equation}

where

\begin{equation*}
\Omega _{N-1}=\frac{2\pi ^{N/2}}{\Gamma (\frac{N}{2})}.
\end{equation*}

We can calculate the quantity $Q_{N}$ which generalises this to $N>3$ black
holes, using Emaparan and Reall (2008), to be

\begin{equation}
MA^{N-2}=c^{2(N-1)}\left[ \frac{(N-2)8\pi G}{(N-1)\Omega _{N-1}}\right]
^{N-2}\left[ \frac{(N-1)\Omega _{N-1}}{16\pi G}\right] ^{N-1}\ .  \label{12}
\end{equation}

For the case of $N=3$ we check that (when the spherical area factor reduces
to $\Omega _{2}=4\pi $)

\begin{equation}
MA=c^{4}\left[ \frac{4\pi G}{\Omega _{N-1}}\right] \times \left[ \frac{%
\Omega _{N-1}}{8\pi G}\right] ^{2}=\frac{c^{4}}{4G}.  \label{13}
\end{equation}%
\ \ 

\subsection{Cosmological evolution}

There is also a cosmological aspect to the maximum force and power
conjectures. Consider the standard Newtonian cosmological picture of an
isotropic and homogeneous universe modelled by an expanding spherical ball
of material with radius proportional to an expansion scale factor $a(t)$. If
we assume $a(t)\propto t^{n}$ then the force generated is proportional to $%
\ddot{a}\propto t^{n-2}$. Thus $F$ will grow as $F=F_{pl}(t/t_{pl})^{n-2}$
for $t>t_{pl}$ if $n>2$. Likewise, the power associated with this expansion
is proportional to $\dot{a}\ddot{a}\ \propto t^{2n-3}$ and grows with time
as $P=P_{pl}(t/t_{pl})^{2n-3}$ when $t>t_{pl\text{ \ }}$under the weaker
condition $n<3/2$; it is constant when $n=3/2$, (Barrow and Cotsakis, 2013) 
\footnote{%
Recently it has been claimed (Dolgov et al, 2014) that the $n=3/2$
expansion, which would be our example of expansion at constant power, gives
a preferred fit to supernovae data in an accelerating zero-curvature
universe in the recent past.}. Here, we see a decoupling of the force and
power conditions. An example of a cosmological evolution with divergent
force is given by the formation of a finite-time sudden singularity in an
isotropic and homogeneous Friedmann universe where $\ddot{a}$ (and the fluid
pressure, $p$) diverges even though $a,\dot{a}$ and the fluid density $\rho $
remain finite, even though $\rho +3p>0$ always.

\section{The effect of the cosmological constant}

Recently, David Thornton (private communication) has raised the question of
how the inclusion of a cosmological constant affects these conclusions about
a maximum force. Recall that Einstein's theory of general relativity states
that matter curves spacetime and spacetime moves matter according to the
Einstein equation 
\begin{equation}
R_{\mu \nu }-{\frac{1}{2}}g^{\alpha \beta }R_{\alpha \beta }\,g_{\mu \nu }=%
\frac{8\pi G}{c^{4}}T_{\mu \nu }-\Lambda g_{\mu \nu },  \label{gr}
\end{equation}%
where the dimensions of the quantities involved are $[x^{\mu }]=L\,,[g_{\mu
\nu }]=1\,,\bigl [R_{\mu \nu }\bigr ]=L^{-2}$ and the energy-momentum tensor 
$T_{\mu \nu }$ has the dimensions of force per unit area, $\bigl [T_{\mu \nu
}\bigl ]=ML^{-1}T^{-2}$ and the cosmological constant $\Lambda $ has
dimensions $[\Lambda ]=L^{-2}$. The inverse of Einstein's constant, $\frac{%
c^{4}}{8\pi G}$, has the dimensions of force $[MLT^{-2}]$, and allows us to
convert from curvature to energy density or stress.

\subsection{New units and bounds}

The cosmological constant $\Lambda $ adds a universal repulsion $\Lambda >0$
(or attraction if $\Lambda <0)$ to the Newtonian gravitational attraction on
a mass $M$ that is proportional to the distance $\mathbf{r}$ (Milne and
McCrea, 1934) 
\begin{equation}
\mathbf{F}_{\Lambda }=M\frac{\Lambda c^{2}}{3}\mathbf{r,}  \label{repulsion}
\end{equation}%
and introduces an additional dimensionful constant into physics. A linear
combination of $\mathbf{F}_{\Lambda }$ and the inverse-square force law of
Newton is the general force law which allows spherical masses to be replaced
by point masses of the same mass (Laplace, 1825, Barrow and Tipler,1986). If 
$\Lambda >0$, as indicated by observations, this may be used to define
another set of fundamental 'de Sitter' units of length, time and mass by
dimensional analysis:

\begin{eqnarray}
L_{ds} &=&\sqrt{\frac{1}{\Lambda }},  \label{ds1} \\
T_{ds} &=&\frac{1}{c}\sqrt{\frac{1}{\Lambda },}  \label{ds2} \\
M_{ds} &=&\frac{\hslash }{c}\sqrt{\Lambda }.  \label{ds3}
\end{eqnarray}

We see that it is not possible to create classical quantities from de Sitter
units that are independent of $\hslash $ if they include $M_{ds\text{ }}$and
so there is no new classical counterpart of eq. (\ref{1}) involving $\Lambda 
$, but we can investigate whether this bound (or one similar) still holds in
the presence of $\Lambda $.

Discussions of the physical significance of the constants of nature often
make use of a mass-radius diagram (Carr and Rees, 1979, Barrow and Tipler,
1986). All bodies, at rest, may be assigned a mass $M$ and a radius or size $%
R$. Since inertial mass, passive gravitational mass, and active
gravitational mass are equal to a high degree of precision, the definition
of mass is unambiguous. The precise definition of radius is not completely
clear (radius of gyration, mean half-diameter, ...), but we will ignore that
small ambiguity here. It follows that all bodies may be assigned a point in
the positive quadrant of the $M-R$ plane.

Large regions of the $M-R$ plane are unoccupied by observable objects
because the Heisenberg Uncertainty Principle gives the lower bound on
observability of real quantum states

\begin{equation}
M\times R>\frac{\hbar }{c},  \label{hup}
\end{equation}%
and the black-hole existence condition gives the constraint \footnote{%
For simplicity we restrict attention to spherically symmetric metrics. For
non-spherically symmetric metrics one might replace $R$ by Thorne's Hoop
radius, (Gibbons, 2009, Cvetic et al, 2011).}

\begin{equation}
R/M>\frac{2G}{c^{2}}.  \label{bh}
\end{equation}

$\ $ Most celestial objects crowd around the lines of constant (atomic or
nuclear) density in the $M-R$ plane where $M\propto R^{3}$.The bounding
rectangular hyperbola (\ref{hup}) and the straight line (\ref{bh}) intersect
close to the point $(M_{pl},R_{pl}).$

One may strengthen these conditions by making more restrictive assumptions
about the body. For example, Buchdahl (1959) obtained the bound 
\begin{equation}
R>\frac{9}{4}\frac{GM}{c^{2}}  \label{buch}
\end{equation}%
for isotropic fluid spheres in the case of vanishing cosmological constant.
Inclusion of the cosmological constant modifies this to, (Mak et al, 2000)

\begin{equation}
R>\frac{2GM}{c^{2}}\frac{1}{\left( 1-\Lambda R^{2}-\frac{1}{9}(1-\frac{%
c^{2}\Lambda R^{3}}{GM})^{2}\right) }\,.  \label{Mak}
\end{equation}

Cosmic repulsion from a positive cosmological constant will blow apart a
body (Hayward et al, 1994, Maeda et al , 1998, Cissoko et al, 1998, Nakao et
al 1991, 1993), unless its representative point lies below the horizontal
line 
\begin{equation}
R=\sqrt{\frac{3}{\Lambda }}.  \label{R}
\end{equation}%
The representative points of all ordinary bodies must lie below the line (%
\ref{R}). and are thus confined inside a triangular region bounded by the
curves (\ref{hup}), (\ref{R})$\ $and (\ref{R}). The intersection of (\ref%
{hup}) and (\ref{R}) gives a lower bound for the mass of any body: 
\begin{equation}
M>\frac{\hbar }{c}\sqrt{\frac{\Lambda }{3}}\,,
\end{equation}%
while the intersection of (\ref{bh}) and (\ref{R}) is gives an upper bound
for the mass of any body: 
\begin{equation}
M<\frac{c^{2}}{G}\sqrt{\frac{3}{\Lambda }}\,.
\end{equation}

\subsection{Kottler-Schwarzschild-de Sitter black holes}

In order to obtain a maximum force estimate in the presence of a
cosmological constant, consider the Kottler-Schwarzschild-de Sitter solution
of Einstein's equations. This is the spherically symmetric vacuum solution
with non-zero cosmological constant: 
\begin{equation}
ds^{2}=-c^{2}\Delta (r)dt^{2}+\frac{dr^{2}}{\Delta (r)}+r^{2}\bigl(d\theta
^{2}+\sin ^{2}\theta d\phi ^{2}\bigr )\,.
\end{equation}%
with

\begin{equation}
\Delta = 1- \frac{2GM}{c^2r} - \frac{1}{3}\Lambda r^2 \,.
\end{equation}

In order to have a static region, $\Delta (r)$ must have two positive roots,
The smaller, at $r=r_{B},$ gives the radius of a black hole event horizon
and the larger, at $r=r_{C},$ of a cosmological event horizon. The
condition\ for two roots is (Gibbons and Hawking, 1977) 
\begin{equation}
3M\sqrt{\Lambda }<\frac{c^{2}}{G}.
\end{equation}

The limiting case was originally found by Nariai (1951) and occurs when 
\begin{equation}
r_{C}=r_{B}=\frac{1}{\sqrt{\Lambda }},
\end{equation}%
and the solution is the metric product $dS_{2}\times S^{2}$

Thus, we see that 
\begin{equation}
\frac{2GM}{c^{2}}\leq r_{B}\leq \frac{1}{\sqrt{\Lambda }}\,,\qquad \frac{1}{%
\sqrt{\Lambda }}\leq r_{C}\leq \sqrt{\frac{3}{\Lambda }}.
\end{equation}

In terms of a force we have 
\begin{equation}
F_{\Lambda }=\frac{Mc^{2}}{3}\Lambda r\,,
\end{equation}%
and so 
\begin{equation}
F_{\Lambda }(r_{B})<\frac{1}{3}Mc^{2}\sqrt{\Lambda }<\frac{c^{4}}{9G}\,.
\end{equation}%
and can be compared with the conjectured maximum force in the absence of a $%
\Lambda $ term given in eq. (\ref{1}).

\section{Conclusions}

We have extended the evidence for the existence of a maximum force, in
general relativity in various ways. We showed how the existence of such a
fundamental bound is linked to the existence of natural units of force that
exclude Planck's constant. Extensions to arbitrary space dimensions reveal a
new quantity that is a candidate for a universal upper bound in general
relativity. We also discussed why a maximum force cannot exist in Newtonian
gravity and how counterexamples describing gravitating with arbitrarily
large forces are avoided by the presence of event horizons. We discussed the
development of strong forces in cosmology and at sudden finite-time
singularities. Finally, we extended the discussion of the maximum force
conjecture to include a cosmological constant. This leads to an additional
system of natural units and new bounds on the maximum and minimum masses of
bodies in the universe. Using the Kottler-Schwarzschild-de Sitter black hole
solution in general relativity we derived a new maximum force bound in the
presence of a positive cosmological constant.

\textbf{Acknowledgements} We would like to thank Christoph Schiller and
David Thornton for stimulating discussions.

\bigskip

\textbf{References}

Barrow J.~D., 1983, Quart. J. Roy. Astron. Soc. 24, 24

Barrow J.~D., 2002, The Constants of Nature, Jonathan Cape, London, chaps.
2-3

Barrow J.D., 2004, Class. Quantum Grav., 21, L79

Barrow J.~D., Cotsakis S., 2013, Phys. Rev. D 88, 067301

Barrow J.D., Tipler F. J., 1986, The Anthropic Cosmological Principle,
Oxford Univ. Press, Oxford, section 5.1, p291.

Buchdahl H.~A., 1959, Phys.\ Rev.\ 116, 1027

Cardoso V., 2013, Gen.\ Rel.\ Grav.\ 45, 2079

Carr B.~J., Rees M.~J., 1979, Nature 278\textbf{,} 605

Cissoko M., Fabris J.~C., Gariel J., Le Denmat G., Santos N.~O., 1998,
gr-qc/9809057

Cvetic M., Gibbons G.W., Pope C.~N., 2011, Class.\ Quant.\ Grav.\ 28, 195001

Dolgov A.D., Halenka V., Tkachev I., 2014, astro-ph 1406.2445

Dyson F., 1963, in Interstellar Communication, ed. Cameron A.G., Benjamin
Inc., New York, chap 12

Emparan R, Reall, H.S., Black Holes in Higher Dimensions, Living Reviews in
Relativity 11, 6 (2008)

Gibbons G.W., 2002, Found.\ Phys.,\ 32, 1891

Gibbons G.W., 2009, gr-qc 0903.1580

Gibbons G.~W., Hawking S.~W., 1977, Phys.\ Rev.\ D 15, 2738

Hayward S., Shiromizu T., Nakao K.-i., 1994, Phys.\ Rev.\ D 49, 5080

Laplace, P.S., 1825, Trait\'{e} de M\'{e}canique C\'{e}leste, Chez J.B.M.
Duprat, Paris,\ Book 2 chap II sec. 12

Maeda K., ~Koike T., Narita M., Ishibashi A., 1998, Phys.\ Rev.\ D 57, 3503

Mak M.~K.,~Dobson Jr. P.~N., Harko T., 2000, Mod.\ Phys.\ Lett.\ A 15, 2153

Marder L., 1959, Proc. Roy. Soc. London A, 252, 45

Milne E.A., McCrea W.H., 1934, Quart. J. Math. 5, 73

Nakao K.-I., Maeda K.-I., Nakamura T., Oohara K.-I., 1991, Phys.\ Rev.\ D
44, 1326

Nakao K.-I., Yamamoto K., Maeda K.-I., 1993, Phys.\ Rev.\ D 47, 3203

Nariai H., 1951, Sci. Rep. Tohoku Univ. Ser. 1, 35, 62

Planck M., 1899, Sitzungsberichte der K\"{o}niglich Preussischen Akademie
der Wissenschaften zu Berlin, 5, 440 and also published as Planck M., 1900,
Ann. d Physik 11, 69 (1900); English translation in M. Planck, 1959, The
Theory of Heat Radiation, transl. M. Masius, Dover, New York (1959).

Schiller C., 1997-2004, Maximum force a simple principle encompassing
general relativity in C. Schiller, Motion Mountain A Hike Beyond Space and
Time Along the Concepts of Modern physics\textit{,}
http://www.motionmountain.net, section 36.

Schiller C., 2006, Int.\ J.\ Theor.\ Phys.,\ 45, 221

Sperhake U., Berti E.,~Cardoso V., 2013, Comptes Rendus Physique, 14, 306

Stoney G.~J., 1881, Phil. Mag. 11, 381

Thomson, W. (Lord Kelvin), Tait P.G., 1879, Treatise on Natural Philosophy,
2nd ed., Cambridge Univ. Press, Cambridge, vol. 1, part 1, p. 227

Vilenkin A., 1981, Phys. Rev. D, 23, 852

Xia Z., 1992, Ann. Math.,135, 411

\end{document}